\documentclass[aps,pre,nofootinbib,showpacs]{revtex4}

\usepackage{amsmath}
\usepackage{graphicx}

\begin{document}

\title{A simple electrostatic model applicable to biomolecular recognition}

\author{T. P. Doerr}
\email{doerr@ncbi.nlm.nih.gov}
\author{Yi-Kuo Yu}
\email{yyu@ncbi.nlm.nih.gov}
\affiliation{National Center for Biotechnology Information,
National Library of Medicine,
National Institutes of Health,
8600 Rockville Pike MSC 6075,
Bethesda, MD 20894-6075}


\begin{abstract}
An exact, analytic solution for a simple electrostatic model 
applicable to biomolecular recognition is presented.
In the model, a layer of high dielectric constant material 
(representative of the solvent, water) whose thickness may vary
separates two regions of low dielectric constant material 
(representative of proteins, DNA, RNA, or similar materials), 
in each of which is embedded a point charge.
For identical charges, the presence of the screening layer always lowers the 
energy compared to the case of point charges in an infinite medium
of low dielectric constant.
Somewhat surprisingly, the presence of a sufficiently thick screening layer 
also lowers the energy compared to the case of point charges
in an infinite medium of {\em high} dielectric constant.
For charges of opposite sign, the screening layer always lowers the energy 
compared to the case of point charges in an infinite medium of either
high or low dielectric constant.
The behavior of the energy leads to a substantially increased repulsive
force between charges of the same sign.
The repulsive force between charges of opposite signs is weaker than
in an infinite medium of low dielectric constant material but stronger
than in an infinite medium of high dielectric constant material.
The presence of this behavior, which we name asymmetric screening, in the
simple system presented here confirms the generality of the behavior 
that was established in a more complicated system of an arbitrary number of
charged dielectric spheres in an infinite solvent.
\end{abstract}


\maketitle

\section{Introduction}

The proper functioning of biomolecular systems depends upon the
aggregation of multiple molecules embedded in a high dielectric constant
solvent (water).
From the medical point of view, there are both normal complexes (such as
ribosomes) and abnormal complexes (such as amyloid formations).
Understanding the microscopic mechanisms involved in the aggregation 
process would illuminate both normal and abnormal states,
and could aid the modification of existing complexes or the design
of new ones.  
This work examines the electrostatic interaction, 
among the most important interactions in biomolecular systems.
\cite{Kauzmann}-\cite{Chaplin2006}

In previous research that developed a scheme for computing to known precision
the energy and forces in a system of an arbitrary number of charged dielectric
spheres embedded in an infinite solvent \cite{tpd06}, 
an effect that was called asymmetric screening was observed.
Namely, the magnitude of attractive electrostatic interactions was decreased 
(relative to point charges in an infinite solvent)
while the magnitude of repulsive electrostatic interactions was increased
(again, relative to point charges in an infinite solvent).
It was speculated that this effect might aid biomolecules such as proteins 
in the adoption of correct conformations and in intermolecular recognition.

This paper presents further studies of this effect in a simplified system that is amenable to complete and thorough analytic examination.  
The simplicity of the model is an advantage in this case because one wishes
to examine in more detail an effect that is already known to occur
in the more general and less symmetric system of spheres mentioned above.
The system studied here can be considered a simplified model of two molecular 
surfaces during the process of binding or aggregation.
Instead of spheres, consider two half-spaces, each with a single point 
charge embedded, separated by an infinite slab of high dielectric constant 
material (water, for example).
If the dielectric constants are swapped, then one would have a model of, for example, a membrane in water.
Separation of variables is used to obatin the potential, and from that the 
energy and the force between the two half-spaces.
It is more convenient to use the surface charge method \cite{tpd06}-\cite{tpd04}
to obtain the density of surface charge induced on the two surfaces.

\section{The General Situation}

\begin{figure} 
\includegraphics{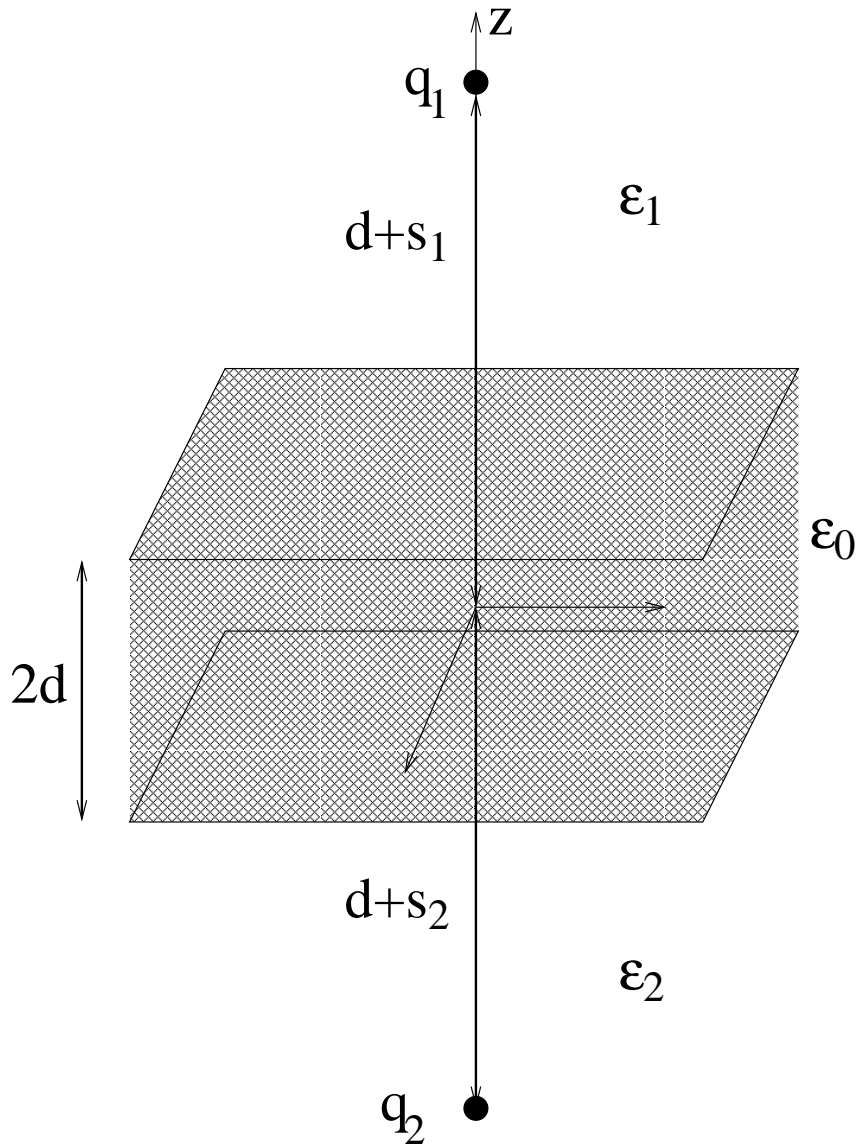}
\caption{\label{system1} The most general situation under consideration.
The shaded region is infinite in the $x$ and $y$ directions, has thickness $2d$
in the $z$ direction, and is filled with a material with dielectric constant 
$\varepsilon_0$.
The origin is chosen so that the distance from the origin to each surface
of the shaded region is $d$.
The unshaded region entirely in the $z>0$ half-space is filled with
a material with dielectric constant $\varepsilon_1$ and contains a charge
$q_1$ on the positive $z$ axis a distance $d+s_1$ from the origin
and a fixed distance $s_1$ from the surface of the shaded region.
The unshaded region entirely in the $z<0$ half-space is filled with
a material with dielectric constant $\varepsilon_2$ and contains a charge
$q_2$ on the negative $z$ axis a distance $d+s_2$ from the origin
and a fixed distance $s_2$ from the surface of the shaded region.}
\end{figure}

Consider a slab of material of thickness $2d$, infinite in the other 
directions, with dielectric constant $\varepsilon_0$ sandwiched between 
two half-spaces filled with materials of dielectric constant
$\varepsilon_1$ and $\varepsilon_2$ respectively.
A charge $q_1$ lies within the external material with dielctric constant
$\varepsilon_1$ a distance $s_1$ from the internal material 
(dielectric constant $\varepsilon_0$);
a charge $q_2$ lies within the other external material 
(with dielctric constant $\varepsilon_2$) a distance $s_2$ from the internal
material and a distance $s_1+s_2+2d$ from the charge $q_1$.
Place the origin of coordinates half way between the two charges.
Place the $z$ axis through the line joining the two charges, perpendicular
to the surfaces of the internal slab of material, and with the positive $z$ 
axis passing through the charge $q_1$, as in Fig.~\ref{system1}.
Because of the symmetry of the system, cylindrical coordinates 
($\rho$, $\phi$, and $z$) will be used.

We wish to find the electric potential ($\Phi$), the electrostatic energy
($U$), and the force ($\vec{F}$) required to pull the external materials apart.
We begin by determining the potential in the general case.
Azimuthal symmetry implies that the potential $\Phi$ is independent of $\phi$.
The symbols $\Phi_0$, $\Phi_1$, and $\Phi_2$ will be used to indicate the
potential in the interior material, in the material entirely in the positive
$z$ region, and in the material entirely in the negative $z$ region 
respectively.
The boundary conditions are
\begin{enumerate}
\item $\Phi \rightarrow 0$ as $z \rightarrow \pm \infty$
\item $\Phi_0(z=d) = \Phi_1(z=d)$
\item $\Phi_2(z=-d) = \Phi_0(z=-d)$
\item $\varepsilon_0\frac{\partial \Phi_0}{\partial z}\left. \right|_{z=d} =
\varepsilon_1\frac{\partial \Phi_1}{\partial z}\left. \right|_{z=d}$
\item $\varepsilon_2\frac{\partial \Phi_2}{\partial z}\left. \right|_{z=-d} =
\varepsilon_0\frac{\partial \Phi_0}{\partial z}\left. \right|_{z=-d}$
\end{enumerate} .

The appropriate general solution of Laplace's equation is
\[
\Phi = \sum_{m=0}^\infty \int_0^\infty J_m(k \rho)(a e^{kz} + b e^{-kz})
(c \sin m\phi + d \cos m\phi) \, \mathrm{d}k 
\rightarrow \int_0^\infty J_0(k \rho)(a e^{kz} + b e^{-kz}) \, \mathrm{d}k ,
\]
because of the azimuthal symmetry.
The appropriate form of the potential of a point charge at $\rho=0$ and $z=z'$
is \cite{jdj}
\[
\frac{1}{\sqrt{\rho^2+(z-z')^2}}=\int_0^\infty e^{-k|z-z'|} J_0(k\rho) 
\, \mathrm{d}k .
\]

The potential in the positive $z$ region of exterior material is a solution
of Laplace's equation plus the potential of the screened point source:
\begin{equation}
\label{p1}
\Phi_1 = \int_0^\infty B_1(k) e^{-kz} J_0(k\rho) \, \mathrm{d}k +
\frac{q_1}{\varepsilon_1}\int_0^\infty e^{-k|z-d-s_1|} J_0(k\rho) 
\, \mathrm{d}k ,
\end{equation}
where boundary condition 1 has deleted one of the exponentials in the 
solution of Laplace's equation.
Similarly, the potential in the negative $z$ region of exterior material is
\begin{equation}
\label{p2}
\Phi_2 = \int_0^\infty A_2(k) e^{kz} J_0(k\rho) \, \mathrm{d}k +
\frac{q_2}{\varepsilon_2}\int_0^\infty e^{-k|z+d+s_2|} J_0(k\rho) 
\, \mathrm{d}k .
\end{equation}
The potential in the interior material is
\begin{equation}
\label{p0}
\Phi_0 = \int_0^\infty (A_0(k)e^{kz} + B_0(k)e^{-kz}) J_0(k\rho) \, \mathrm{d}k .
\end{equation}
Boundary conditions 2-5 determine the coefficients:
\begin{subequations}
\label{4by4soln}
\begin{eqnarray}
B_1(k) & = & e^{k(d-s_1-s_2)} \frac{e^{ks_2}(\varepsilon_0+\varepsilon_1)
(\varepsilon_0-\varepsilon_2)q_1 - e^{k(4d+s_2)}(\varepsilon_0-\varepsilon_1)
(\varepsilon_0+\varepsilon_2)q_1 + 4e^{k(2d+s_1)}\varepsilon_0\varepsilon_1q_2}
{-(\varepsilon_0-\varepsilon_1)\varepsilon_1(\varepsilon_0-\varepsilon_2)+
e^{4kd}\varepsilon_1(\varepsilon_0+\varepsilon_1)(\varepsilon_0+\varepsilon_2)}
\\
A_0(k) & = & 2 e^{k(d-s_1-s_2)} \frac{e^{k(2d+s_2)}(\varepsilon_0+\varepsilon_2)q_1 + e^{ks_1}(\varepsilon_0-\varepsilon_1)q_2}{-(\varepsilon_0-\varepsilon_1)
(\varepsilon_0-\varepsilon_2) + e^{4kd}(\varepsilon_0+\varepsilon_1)
(\varepsilon_0+\varepsilon_2)}  \\
B_0(k) & = & 2 e^{k(d-s_1-s_2)} \frac{e^{ks_2}(\varepsilon_0-\varepsilon_2)q_1+
e^{k(2d+s_1)}(\varepsilon_0+\varepsilon_1)q_2}{-(\varepsilon_0-\varepsilon_1)
(\varepsilon_0-\varepsilon_2) + e^{4kd}(\varepsilon_0+\varepsilon_1)
(\varepsilon_0+\varepsilon_2)} \\
A_2(k) & = & e^{k(d-s_1-s_2)} \frac{4e^{k(2d+s_2)}\varepsilon_0\varepsilon_2q_1-e^{k(4d+s_1)}(\varepsilon_0+\varepsilon_1)(\varepsilon_0-\varepsilon_2)q_2+
e^{ks_1}(\varepsilon_0-\varepsilon_1)(\varepsilon_0+\varepsilon_2)q_2}
{-(\varepsilon_0-\varepsilon_1)\varepsilon_2(\varepsilon_0-\varepsilon_2)+
e^{4kd}\varepsilon_2(\varepsilon_0+\varepsilon_1)(\varepsilon_0+\varepsilon_2)}.
\end{eqnarray}
\end{subequations}
Not surprisingly, interchanging the indices 1 and 2 in the expression for $B_1$
turns it into $A_2$.

The distribution of free charge (the two point charges) and the potential
determine the energy:
\begin{equation}
\label{e1}
U = \frac{1}{2} \int \rho_f \Phi = 
\frac{q_1}{2} \Phi'_1(\rho=0, z=d+s_1) + \frac{q_2}{2} \Phi'_2(\rho=0, z=-d-s_2) ,
\end{equation}
where the primes on the potentials indicate that the potential of the point
charge in the corresponding region has been subtracted out in order to avoid infinite self-energies.
Substitution of Eq.~(\ref{p1}), Eq.~(\ref{p2}), and Eq.~(\ref{4by4soln}) 
into Eq.~(\ref{e1}) yields
\begin{eqnarray}
\label{genenergy}
U & = & \frac{4 q_1 q_2 \varepsilon_0}
{(\varepsilon_0+\varepsilon_1)(\varepsilon_0+\varepsilon_2)}
\int_0^\infty \frac{e^{-k(2d+s_1+s_2)}} {1 - \alpha_1 \alpha_2 e^{-4kd}} 
\, \mathrm{d}k +
\frac{q_1^2}{2 \varepsilon_1} \int_0^\infty
\frac{e^{-2ks_1}(e^{-4kd}\alpha_2 - \alpha_1)}{1 - \alpha_1 \alpha_2 e^{-4kd}} 
\, \mathrm{d}k \nonumber \\
& & + \frac{q_2^2}{2 \varepsilon_2} \int_0^\infty
\frac{e^{-2ks_2}(e^{-4kd}\alpha_1 - \alpha_2)}{1 - \alpha_1 \alpha_2 e^{-4kd}} 
\, \mathrm{d}k ,
\end{eqnarray}
where 
$\alpha_1 \equiv (\varepsilon_0-\varepsilon_1)/(\varepsilon_0+\varepsilon_1)$ 
and 
$\alpha_2 \equiv (\varepsilon_0-\varepsilon_2)/(\varepsilon_0+\varepsilon_2)$.

Because we imagine this situation to be a simplified model of two molecular 
surfaces separated by a layer of water, the force should be obtained by
imagining that the charges are fixed with respect to the materials in which
they are embedded, but the thickness of the interior slab is allowed to vary.
In other words, the force we are considering is the negative of the derivative 
of the energy with respect to $2d$:  
\[
\vec{F}=-\frac{\partial U}{\partial (2d)}\hat{z} ,
\]
or in scalar form for the magnitude
\[
F = -\frac{1}{2}\frac{\partial U}{\partial d} .
\]
Clearly, this simple model neglects any internal rearrangement of the molecules 
during the process of interaction, an effect that is believed to be important
in many cases.
However, while a model designed to capture the behavior of specific molecules
would need to include such an effect, our purpose is only to 
investigate one particular interaction, the very important electrostatic 
interaction, and so this point is not a concern here.
The force is
\begin{eqnarray}
\label{genforce}
F & = & \frac{4 q_1 q_2 \varepsilon_0}
{(\varepsilon_0+\varepsilon_1)(\varepsilon_0+\varepsilon_2)}
\int_0^\infty e^{-k(2d+s_1+s_2)} k 
\frac{1+\alpha_1 \alpha_2 e^{-4kd}}{(1-\alpha_1 \alpha_2 e^{-4kd})^2}
\, \mathrm{d}k \nonumber \\
& & + \frac{q_1^2}{\varepsilon_1}\alpha_2(1-\alpha_1^2) \int_0^\infty
\frac{e^{-k(2s_1+4d)} k}{(1-\alpha_1 \alpha_2 e^{-4kd})^2} \, \mathrm{d}k 
\nonumber \\
& & + \frac{q_2^2}{\varepsilon_2}\alpha_1(1-\alpha_2^2) \int_0^\infty
\frac{e^{-k(2s_2+4d)} k}{(1-\alpha_1 \alpha_2 e^{-4kd})^2} \, \mathrm{d}k .
\end{eqnarray}
We now examine two particular cases.

\section{Two Identical Charges in Identical Media}\label{lc}

\begin{figure} 
\includegraphics{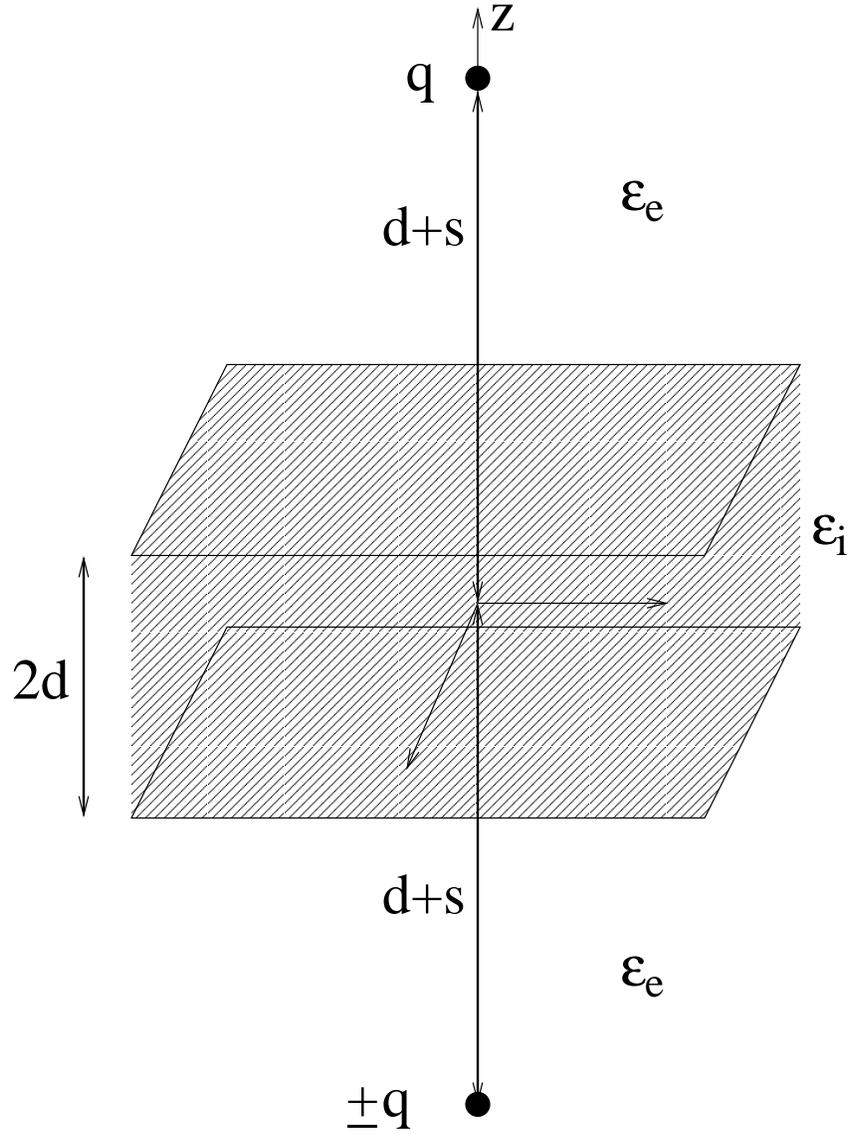}
\caption{\label{system2} A simplified situation considered in detail.
The charges are now of equal magnitute and are constrained to be the same 
distance from the origin.
The cases of identical charges and of opposite charges are both considered.
Both unshaded regions have the same dielectric constant, referred to 
as $\varepsilon_\mathrm{e}$.
The dielectric constant of the shaded slab is now referred to as 
$\varepsilon_\mathrm{i}$.}
\end{figure}

Let $q_1 = q_2 \equiv q$, 
$\varepsilon_1 = \varepsilon_2 \equiv \varepsilon_\mathrm{e}$,
$\varepsilon_0 \equiv \varepsilon_\mathrm{i}$, and $s_1 = s_2 \equiv s$.
We are now considering
a slab of material (thickness $2d$ and infinite in the other directions) with
dielectric constant $\varepsilon_\mathrm{i}$ sandwiched between two half-spaces
filled with a material of dielctric constant $\varepsilon_\mathrm{e}$.
(Internal material is indicated by the subscript `i', and external material
is indicated by subscript `e'.)
A charge $q$ lies in the external material a distance $s$ from the internal 
material.
An identical charge $q$ lies in the other semi-infinite external material a 
distance $s$ from the internal material and a distance $2s+2d$ from the 
other charge.  
See Fig.~\ref{system2}, with the positive charge chosen.
The potential, the energy, and the force follow upon making the appropriate
substitutions in Eqs.~(\ref{p1}-\ref{p0}), Eq.~(\ref{genenergy}), 
and Eq.~(\ref{genforce}) respectively.
(Alternatively, it is a simple matter to set up and solve the boundary value 
problem for this particular situation.)

Making the appropriate substitutions in Eq.~(\ref{genenergy}), letting 
$\alpha = (\varepsilon_\mathrm{i}-\varepsilon_\mathrm{e})/
(\varepsilon_\mathrm{i}+\varepsilon_\mathrm{e})$,
and using the identity
$(4\varepsilon_\mathrm{i}\varepsilon_\mathrm{e}/
(\varepsilon_\mathrm{i}+\varepsilon_\mathrm{e})^2) = 1 - \alpha^2$,
one finds the energy:
\begin{eqnarray}
\label{lcu}
U & = & \frac{q^2}{\varepsilon_\mathrm{e}} \int_0^\infty e^{-2ks}
\frac{e^{-2kd}(4\varepsilon_\mathrm{i}\varepsilon_\mathrm{e}/
(\varepsilon_\mathrm{i}-\varepsilon_\mathrm{e})^2) +
\alpha(e^{-4kd}-1)}{1-\alpha^2e^{-4kd}} \, \mathrm{d}k \nonumber \\ 
  & = & \frac{q^2}{\varepsilon_\mathrm{e}} \int_0^\infty e^{-2k(s+d)}
\frac{1-\alpha e^{2kd}}{1-\alpha e^{-2kd}} \, \mathrm{d}k .
\end{eqnarray}
One may evaluate the integral by expanding the denominator in a series:
\begin{eqnarray}
\label{lcus}
U & = & \frac{q^2}{\varepsilon_\mathrm{e}} \int_0^\infty \sum_{n=0}^\infty
\left( \alpha^n e^{-2k(s+(n+1)d)} -
\alpha^{n+1} e^{-2k(s+nd)}\right) \, \mathrm{d}k \nonumber \\
  & = & \frac{q^2}{2 \varepsilon_\mathrm{e}} \sum_{n=0}^\infty
\left( \frac{\alpha^n}{s+(n+1)d} - \frac{\alpha^{n+1}}{s+nd} \right) \nonumber\\
  & = & \frac{q^2 (1-\alpha^2)}{2 \varepsilon_\mathrm{e} \alpha} 
\sum_{n=0}^\infty \frac{\alpha^n}{s+nd} -
\frac{q^2}{2 \varepsilon_\mathrm{e} \alpha s} \nonumber \\
  & = & \frac{q^2(1-\alpha^2)}{2\varepsilon_\mathrm{e} \alpha s} 
{}_2F_1\left(\frac{s}{d}, 1; \frac{s}{d}+1; \alpha\right) -
\frac{q^2}{2\varepsilon_\mathrm{e} \alpha s} .
\end{eqnarray}
where ${}_2F_1$ is a Gauss hypergeometric function.

Even though the series in Eq.~(\ref{lcus}) was obtained by separation of
variables, it can be interpreted as the effect of an infinite sequence of
image charges.  
The charges have separations $2s+2nd$ for $n=0,1,2,\ldots$.
The magnitude of the image charges can be read off from the coefficients of
$q/(\varepsilon_\mathrm{e}(2s+2nd))$ with appropriate care taken to 
separate out the direct interaction of the free charges.
This interpretation brings to mind recent work that used an approximate 
series of image charges to study a pair of membranes in a solvent
of water and ions.\cite{Pincus2008}

\begin{figure} 
\includegraphics[width=5in]{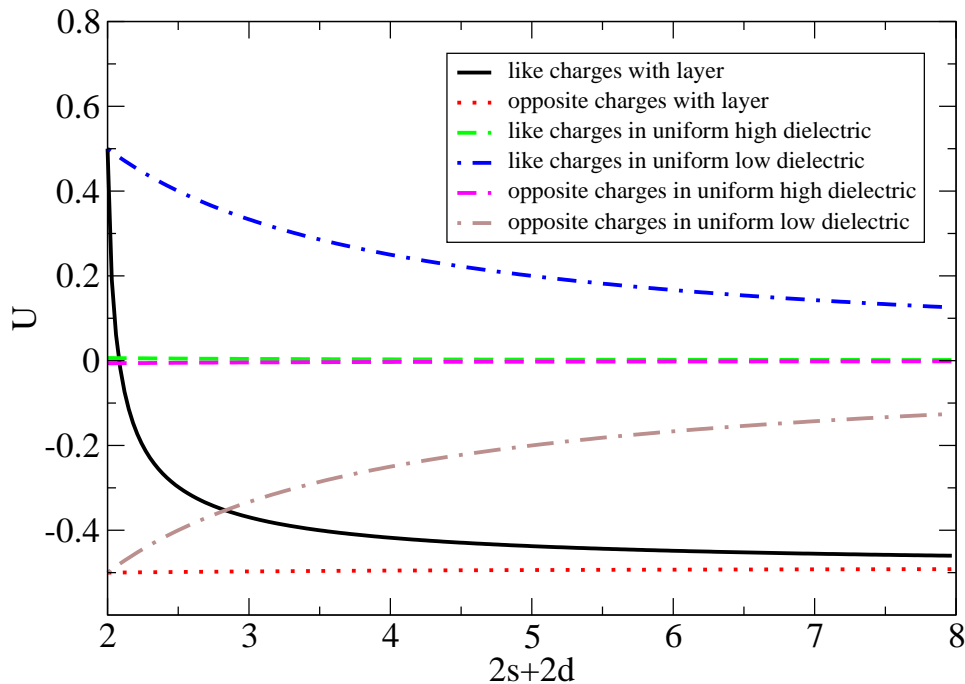}
\caption{\label{U_d} Graphs of the energy as a function of separation, both for
identical charges and for opposite charges.
For comparison, the energy of point charges, both identical and opposite,
in an infinite uniform medium (both $\varepsilon_\mathrm{e}$ and
$\varepsilon_\mathrm{i}$) is shown.
The calculations are for $\varepsilon_\mathrm{e}=1$, 
$\varepsilon_\mathrm{i}=80$, $s=1$, and $q=1$.
For opposite charges separated by a high dielectric layer, 
the energy varies little.
For like charges separated by a high dielectric layer, 
the energy at small separations changes rapidly.}
\end{figure}

\begin{figure} 
\includegraphics[width=5in]{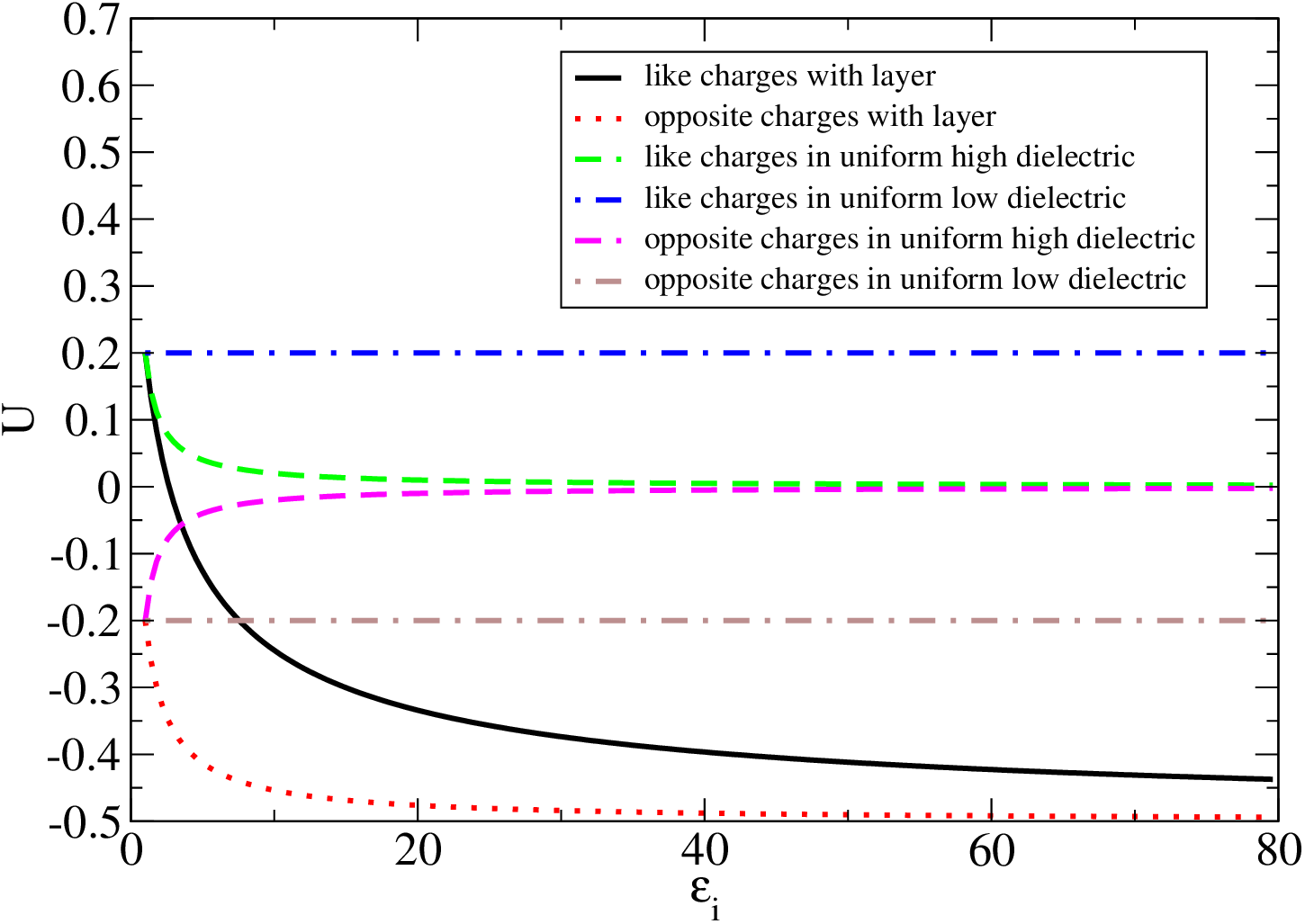}
\caption{\label{U_epse} Graphs of the energy as a function of 
$\varepsilon_\mathrm{i}$, both for
identical charges and for opposite charges.
For comparison, the energy of point charges, both identical and opposite,
in an infinite uniform medium (both $\varepsilon_\mathrm{e}$ and
$\varepsilon_\mathrm{i}$) is shown.
The calculations are for $2s+2d=5$,
$\varepsilon_\mathrm{e}=1$, $s=1$, and $q=1$.}
\end{figure}

Because the dielectric constant of water ($\approx 80$\cite{crc}) is much
larger than the dielectric constant of protein ($\approx 4$\cite{honig86}), 
we are most interested in screening situation: $0 \le \alpha \le 1$.
In the limit $\alpha \rightarrow 1$, the interior slab becomes metallic.
In this case we find that $U=-q^2/(\varepsilon_\mathrm{e} 2 s)$, 
which is just the
interaction energy of each free charge with its image charge due to the metal;
the two free charges do not `feel' each other.
If the media all have the same dielectric constant, then $\alpha = 0$ and
$U=q^2/(\varepsilon_\mathrm{e} 2 (s+d))$, 
which is simply the energy of two charges
in an infinite dielectric medium.
Similarly, if $d=0$ we find the obvious result 
$U=q^2/(\varepsilon_\mathrm{e} 2 s)$.
Finally, in the limit that $d \rightarrow \infty$, 
$U \rightarrow -(q^2 \alpha)/(\varepsilon_\mathrm{e} 2 s) < 0$.
In this case, the two fixed charges do not see each other, but each point
charge can still induce a charge density on the nearby surface, and this 
process will always reduce the energy. 
Therefore $U$ is negative in this limit.
The behavior just summarized can be seen in Fig.~\ref{U_d} and Fig.~\ref{U_epse}.

\begin{figure} 
\includegraphics[width=5in]{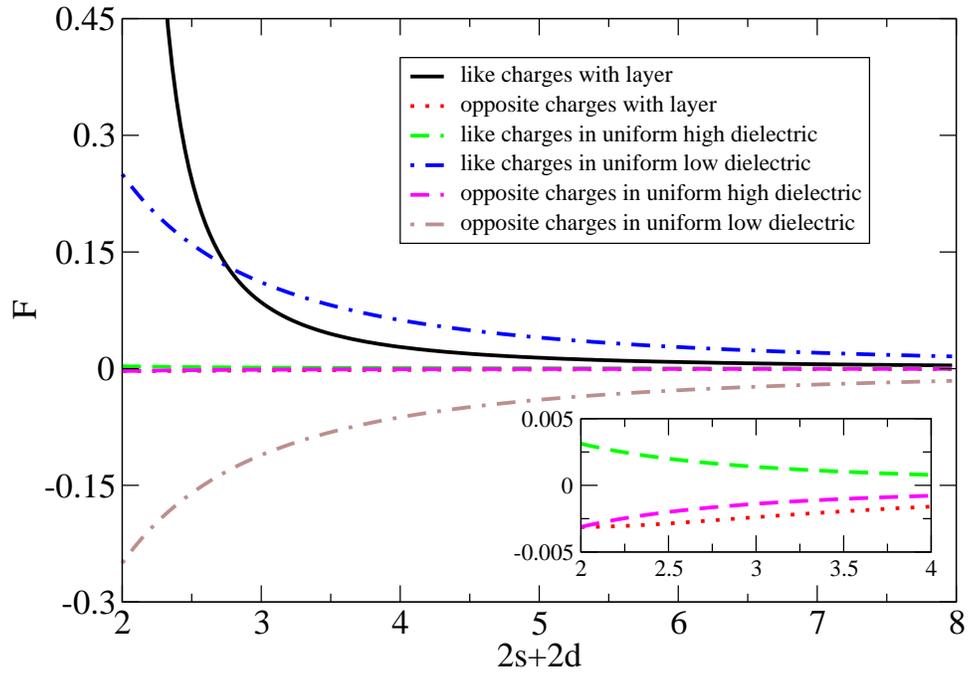}
\caption{\label{F_d} Graphs of the force as a function of separation, both for
identical charges and for opposite charges.
For comparison, the force between point charges, both identical and opposite,
in an infinite uniform medium (both $\varepsilon_\mathrm{e}$ and
$\varepsilon_\mathrm{i}$) is shown.
The calculations are for $\varepsilon_\mathrm{e}=1$, 
$\varepsilon_\mathrm{i}=80$, $s=1$, and $q=1$.
The inset is a close-up of the three curves near the $x$ axis for small 
separations.}
\end{figure}

\begin{figure} 
\includegraphics[width=5in]{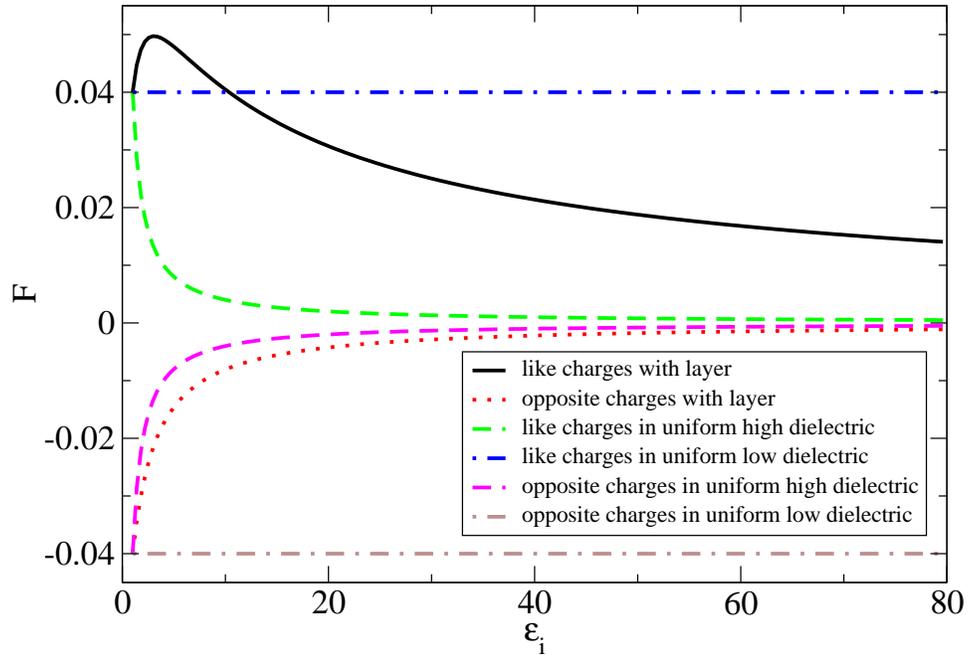}
\caption{\label{F_epse} Graphs of the force as a function of 
$\varepsilon_\mathrm{i}$, both for
identical charges and for opposite charges.
For comparison, the force between point charges, both identical and opposite,
in an infinite uniform medium (both $\varepsilon_\mathrm{e}$ and
$\varepsilon_\mathrm{i}$) is shown.
The calculations are for $2s+2d=5$
$\varepsilon_\mathrm{e}=1$, $s=1$, and $q=1$.}
\end{figure}

Making the appropriate substitutions
in Eq.~(\ref{genforce}) and again using the identity
$(4\varepsilon_\mathrm{i}\varepsilon_\mathrm{e}/
(\varepsilon_\mathrm{i}+\varepsilon_\mathrm{e})^2) = 1 - \alpha^2$,
one finds the force:
\begin{equation}
\label{lcf}
F = \frac{q^2(1-\alpha^2)}{\varepsilon_\mathrm{e}} \int_0^\infty
\frac{k e^{-2k(d+s)}}{(1-\alpha e^{-2kd})^2} \, \mathrm{d}k .
\end{equation}
Rather than performing a similar procedure with series to evaluate the integral,
one may simply differentiate the series for $U$:
\begin{eqnarray}
\label{lcfs}
F & = & -\frac{q^2 \alpha (1-\alpha^{-2})}{\varepsilon_\mathrm{e}} \sum_{n=0}^\infty
\frac{n \alpha^n}{(2s+n2d)^2} \nonumber \\
& = & \frac{q^2 (1-\alpha^2)}{4 \varepsilon_\mathrm{e} \alpha} 
\sum_{n=0}^\infty \frac{n \alpha^n}{(s+nd)^2} .
\end{eqnarray}
As noted above, for the case of complete screening ({\em i.e.}, $\alpha=1$)
the free charges do not `feel' each other.  
As expected, the force vanishes in this case.
If the media all have the same dielectric constant, then $\alpha = 0$ and
$F=q^2/(\varepsilon_\mathrm{e} (2s+2d)^2)$, 
the force between two identical charges
in an infinite dielectric medium.
On the other hand, if $d=0$ we find the curious result 
$F=(q^2 \varepsilon_\mathrm{i})/(\varepsilon_\mathrm{e}^2 4 s^2)$.
When $d=0$ one might expect $F$ not to depend on $\varepsilon_\mathrm{i}$.
However, $F(d)$ samples $U(d)$ in the vicinity of $d$, and even when $d=0$
a dependence is generated on $\varepsilon_\mathrm{i}$, which characterizes
the material that would fill the gap if one were to draw the two outer 
regions apart.
Indeed, for $d=0$ and $\varepsilon_\mathrm{i} \rightarrow 1$, the force becomes
infinite, {\em i.e.}, the energy changes discontinuously 
at $d=0$ if $\alpha=1$.
The behavior just summarized can be seen in Fig.~\ref{F_d} and Fig.~\ref{F_epse}.

The difference between $U$ and the energy of two point charges in an infinite
medium of dielectric constant $\varepsilon_\mathrm{e}$ is defined to be 
$\Delta U$.
(This could not be calculated in the general case because in that case
there is no single exterior material.)
One finds
\begin{equation}
\label{lcdu}
\Delta U = \frac{q^2}{\varepsilon_\mathrm{e}} \int_0^\infty e^{-2k(s+d)}
\left(\frac{1 - \alpha e^{2kd}}{1 - \alpha e^{-2kd}} - 1 \right) \, \mathrm{d}k
= -\frac{q^2 \alpha}{\varepsilon_\mathrm{e}} \int_0^\infty e^{-2ks}
\frac{1 - e^{-4kd}}{1 -\alpha e^{-2kd}} \, \mathrm{d}k .
\end{equation}
Notice that $\Delta U \le 0$ in the case of screening ($\alpha > 0$),
which makes sense because the energy should be lowered by replacing a portion
of the low dielectric constant material 
with higher dielectric constant material.
If $\alpha=0$, the energy $U$ is the same as the term we have just subtracted
off, so $\Delta U=0$.
Similarly, if $d=0$, then $\Delta U=0$.

The force difference $\Delta F$ corresponding to $\Delta U$ can be obtained
either from the expression for $\Delta U$ or the expression for $F$:
\begin{equation}
\label{lcdf}
\Delta F = \frac{q^2}{\varepsilon_\mathrm{e}} \int_0^\infty k e^{-2k(d+s)} 
\left( \frac{(1-\alpha^2)}{(1-\alpha e^{-2kd})^2} - 1 \right) \, \mathrm{d}k .
\end{equation}
In the case of $d=0$ we find that
$\Delta F=\frac{q^2 \alpha}{2 \varepsilon_\mathrm{e} s^2 (1-\alpha)}$.
If $\alpha=1$, then 
$\Delta F=-q^2/(\varepsilon_\mathrm{e}(2s+2d)^2)<0$ which,
as expected, is just the term we subtracted off to form $\Delta F$.
Clearly $\Delta F=0$ if $\alpha=0$.
The behavior of $\Delta F$ for small but non-zero $\alpha$ may be 
deduced from the series expression for $F$:
\begin{eqnarray*}
\Delta F & = & 
\frac{q^2}{4 \varepsilon_\mathrm{e}} \sum_{n=1}^\infty
\frac{n (1-\alpha^2) \alpha^{n}}{\alpha(s+nd)^2} 
- \frac{q^2}{4 \varepsilon_\mathrm{e}(s+d)^2} \\
& > & \frac{q^2}{4 \varepsilon_\mathrm{e}} \sum_{n=1}^\infty
\frac{n (1-\alpha^2) \alpha^{n}}{\alpha (ns+nd)^2}
- \frac{q^2}{4 \varepsilon_\mathrm{e}(s+d)^2} \\
& = & \frac{q^2}{4 \varepsilon_\mathrm{e} (s+d)^2} \frac{(1-\alpha^2)}{\alpha} 
\sum_{n=1}^\infty \frac{\alpha^n}{n}
- \frac{q^2}{4 \varepsilon_\mathrm{e}(s+d)^2} \\
& = & \frac{q^2}{4 \varepsilon_\mathrm{e} (s+d)^2} \frac{\alpha}{2} 
+ {\cal O}(\alpha^2).
\end{eqnarray*}
When $\Delta F > 0$, the repulsion between identical charges is stronger than
the case when both identical charges are in one uniform medium with dielectric
constant $\varepsilon_\mathrm{e}$.  
Upon letting $\varepsilon_\mathrm{e} \rightarrow 1$ 
(see Fig.~\ref{F_d} and Fig.~\ref{F_epse}),
we see that one can have a repulsion larger than in vacuum,
a counter-intuitive conclusion.
The origin of this behavior can be deduced by returning to 
Eq.~(\ref{genenergy}),
the energy for the more general situation first described.
Setting $q_1=0$, $q_2=q$, and $s_2=s$
but retaining distinct dielectric constants in each region,
we find
\begin{eqnarray*}
U & = & \frac{q^2}{2 \varepsilon_2} \int_0^\infty
\frac{e^{-2ks}(e^{-4kd}\alpha_1 - \alpha_2)}{1 - \alpha_1 \alpha_2 e^{-4kd}} 
\, \mathrm{d}k \\
& = & \frac{q^2}{2 \varepsilon_2} \sum_{n=0}^\infty \alpha_1^n \alpha_2^n
\left[ \frac{\alpha_1}{2s+4(n+1)d} - \frac{\alpha_2}{2s+4nd} \right] ,
\end{eqnarray*}
and 
\[
F = \frac{q^2}{\varepsilon_2} \sum_{n=0}^\infty \alpha_1^n \alpha_2^n
\left[ \frac{(n+1)\alpha_1}{(2s+4(n+1)d)^2} - 
\frac{n\alpha_2}{(2s+4nd)^2} \right] .
\]
Each factor of $\alpha_1$ ($\alpha_2$) indicates an image reflection 
across the surface of the material with dielectric constant
$\varepsilon_1$ ($\varepsilon_2$).
Notice that the induced charge of the leading term (proportional to 
$\alpha_1$) is the same sign as the free charge because the image
charge is located on the low dielectric side of the interface.
If the image charge were located on the high dielectric side of the
interface ($\varepsilon_0<\varepsilon_1$ and $\varepsilon_0<\varepsilon_2$) 
then the induced charge would have the opposite sign leading to an
attractive force similar to the more familiar case of a charge
near a conductor.

Now consider the energy and force differences 
($\widetilde{\Delta U}$ and $\widetilde{\Delta F}$) when the comparison is made
to the interaction with the $\varepsilon_\mathrm{i}$ material everywhere.
The energy difference in this case, $\widetilde{\Delta U}$, is
\begin{eqnarray}
\label{lcdut}
\widetilde{\Delta U} & = & \frac{q^2}{\varepsilon_\mathrm{e}} 
\int_0^\infty e^{-2k(s+d)}
\left(\frac{1 - \alpha e^{2kd}}{1 - \alpha e^{-2kd}} - 
\frac{\varepsilon_\mathrm{e}}{\varepsilon_\mathrm{i}} \right) 
\, \mathrm{d}k \nonumber \\
& = & \frac{q^2}{\varepsilon_\mathrm{i}\varepsilon_\mathrm{e}} 
\int_0^\infty e^{-2k(s+d)}
\frac{(\varepsilon_\mathrm{i}-\varepsilon_\mathrm{e}) + 
\alpha (\varepsilon_\mathrm{e} e^{-2kd} - \varepsilon_\mathrm{i} e^{2kd})}
{1-\alpha e^{-2kd}} \, \mathrm{d}k .
\end{eqnarray}
In order to understand the behavior of $\widetilde{\Delta U}$, we observe that
$\widetilde{\Delta U}(d=0) = 
(q^2(\varepsilon_\mathrm{i}-\varepsilon_\mathrm{e}))/(\varepsilon_\mathrm{i}\varepsilon_\mathrm{e}2s) \ge 0$ with equality
when $\varepsilon_\mathrm{i}=\varepsilon_\mathrm{e}$ ({\em i.e.}, $\alpha=0$).
However, as $d\rightarrow\infty$, 
$\widetilde{\Delta U}\rightarrow -(q^2\alpha)/(2\varepsilon_\mathrm{e}s) \le 0$.
Evidently, for any positive $\alpha$, $\widetilde{\Delta U}$ is positive for
small $d$ and becomes negative for sufficiently large $d$.
This behavior can be inferred from Fig.~\ref{U_d}.
Given that $(\varepsilon_\mathrm{i}/\varepsilon_\mathrm{e})(1-\alpha)=(1+\alpha$), 
$\widetilde{\Delta F}$ is
\begin{eqnarray}
\label{lcdft}
\widetilde{\Delta F} & = &  
\frac{q^2}{\varepsilon_\mathrm{i}} \int_0^\infty k e^{-2k(s+d)}
\left[ \frac{(\varepsilon_\mathrm{i}/\varepsilon_\mathrm{e})(1-\alpha^2)}
{(1-\alpha e^{-2kd})^2} - 1 \right] \, \mathrm{d}k \nonumber \\
& = & \frac{q^2}{\varepsilon_\mathrm{i}} \int_0^\infty k e^{-2k(s+d)}
\left[ \frac{(1+\alpha)^2}{(1-\alpha e^{-2kd})^2} - 1 \right] \, \mathrm{d}k \\
& \ge & \frac{q^2}{\varepsilon_\mathrm{i}} \int_0^\infty k e^{-2k(s+d)}
[(1+\alpha)^2 - 1] \, \mathrm{d}k \ge 0 \nonumber ,
\end{eqnarray}
which guarantees that $\widetilde{\Delta F} \ge 0$, as would be expected
based upon Figs.~\ref{F_d} and \ref{F_epse}.

\section{Two Opposite Charges in Identical Media}\label{oc}

Consider the same situation as in the previous section 
except that the two charges are of opposite sign.
Namely, let $q_1 \equiv q$, $q_2 \equiv -q$, 
$\varepsilon_1 = \varepsilon_2 \equiv \varepsilon_\mathrm{e}$,
$\varepsilon_0 \equiv \varepsilon_\mathrm{i}$, and $s_1 = s_2 \equiv s$.
See Fig.~\ref{system2}, with the negative charge chosen.
The potential, the energy, and the force follow upon making the appropriate
substitutions in Eqs.~(\ref{p1}-\ref{p0}), Eq.~(\ref{genenergy}), 
and Eq.~(\ref{genforce}) respectively.
(Alternatively, it is a simply matter to set up and solve the boundary value 
problem for this particular situation.)

Making the appropriate substitutions in Eq.~(\ref{genenergy}), letting 
$\alpha = (\varepsilon_\mathrm{i}-\varepsilon_\mathrm{e})/
(\varepsilon_\mathrm{i}+\varepsilon_\mathrm{e})$,
and using the identity
$(4\varepsilon_\mathrm{i}\varepsilon_\mathrm{e}/(\varepsilon_\mathrm{i}-\varepsilon_\mathrm{e})^2) = 1 - \alpha^2$,
one finds the energy:
\begin{eqnarray}
\label{ocu}
U & = & -\frac{q^2}{\varepsilon_\mathrm{e}} \int_0^\infty e^{-2ks}
\frac{e^{-2kd}(1-\alpha^2) - \alpha(e^{-4kd}-1)}{1-\alpha^2e^{-4kd}}
\, \mathrm{d}k \nonumber \\
  & = & -\frac{q^2}{\varepsilon_\mathrm{e}} \int_0^\infty e^{-2k(s+d)}
\frac{1+\alpha e^{2kd}}{1+\alpha e^{-2kd}} \, \mathrm{d}k .
\end{eqnarray}
Again, we are most interested in screening situation: $0 \le \alpha \le 1$.
When $\alpha=1$ (perfect screening), 
we find that $U=-q^2/(\varepsilon_\mathrm{e} 2 s)$, which is just the
interaction energy of each free charge with its image charge due to the metal;
the two free charges do not `feel' each other.
If the media all have the same dielectric constant, then $\alpha = 0$ and
$U=-q^2/(\varepsilon_\mathrm{e} 2 (s+d))$, which is simply the energy of two charges
in an infinite dielectric medium.
Similarly, if $d=0$ we find the obvious result $U=-q^2/(\varepsilon_\mathrm{e} 2 s)$.
Finally, in the limit that $d \rightarrow \infty$,
$U \rightarrow -(q^2 \alpha)/(\varepsilon_\mathrm{e} 2 s) < 0$.
In this case, the two fixed charges do not see each other, but each point charge can still induce a charge density on the nearby surface, and this
process will always reduce the energy.
Note that if $\alpha$ is close to unity ({\em e.g.}, a water solvent),
$U$ varies little as $d$ goes from 0 to $\infty$.
The behavior just summarized can be seen in Fig.~\ref{U_d} and Fig.~\ref{U_epse}.
Comparing Eq.~(\ref{ocu}) with Eq.~(\ref{lcu}), 
one sees that the series for $U$ is the series for identical charges with 
an overall minus sign and the substitution $\alpha \rightarrow -\alpha$.

Making the appropriate substitutions
in Eq.~(\ref{genforce}) and again using the identity
$(4\varepsilon_\mathrm{i}\varepsilon_\mathrm{e}/(\varepsilon_\mathrm{i}-\varepsilon_\mathrm{e})^2) = 1 - \alpha^2$,
one finds the force:
\begin{equation}
\label{ocf}
F = -\frac{q^2(1-\alpha^2)}{\varepsilon_\mathrm{e}} \int_0^\infty
\frac{k e^{-2k(d+s)}}{(1+\alpha e^{-2kd})^2} \, \mathrm{d}k .
\end{equation}
As noted above, for the case of complete screening ({\em i.e.}, $\alpha=1$)
the free charges do not `feel' each other.  
As expected, the force vanishes in this case.
If the media all have the same dielectric constant, then $\alpha = 0$ and
$F=-q^2/(\varepsilon_\mathrm{e} (2s+2d)^2)$, the force between two opposite charges
in an infinite dielectric medium.
If $d=0$ we find the somewhat non-obvious result $F=-q^2/(\varepsilon_\mathrm{i} 4 s^2)$,
the explanation for which is the same as in the case of identical charges.
The behavior of the force in the case of opposite charges is more consistent
with naive intuition:  the force with a high dielectric layer is somewhere
in between the force with low dielectric everywhere and the force with
high dielectric everywhere.
The behavior just summarized can be seen in Fig.~\ref{F_d} and Fig.~\ref{F_epse}.
Comparing Eq.~(\ref{ocf}) with Eq.~(\ref{lcf}), 
one sees that the series for $F$ is the series for identical charges with 
an overall minus sign and the substitution $\alpha \rightarrow -\alpha$.

The energy difference $\Delta U$ is now calculated along the lines used in 
the case of identical charges:
\begin{eqnarray}
\label{ocdu}
\Delta U & = & \frac{q^2}{\varepsilon_\mathrm{e}} \int_0^\infty e^{-2k(s+d)}
\left[ 1 - \frac{1+\alpha e^{2kd}}{1+\alpha e^{-2kd}} \right] \, \mathrm{d}k \\
& = & - \frac{2 q^2 \alpha}{\varepsilon_\mathrm{e}} \int_0^\infty e^{-2k(s+d)}
\frac{\sinh 2kd}{1+\alpha e^{-2kd}} \, \mathrm{d}k \le 0 \nonumber .
\end{eqnarray}
Since $U(d=0)=-q^2/(\varepsilon_\mathrm{e}(2s+2d))$ and 
$U \rightarrow -q^2\alpha/(2\varepsilon_\mathrm{e}s)$ as $d \rightarrow \infty$,
it is clear that $\Delta U$ should be negative (see Fig.~\ref{U_d}).
As expected, the energy difference $\Delta U$ vanishes both for 
$d=0$ and for $\alpha=0$.
For $\alpha=1$, each charge interacts with its image charge, 
and therefore $\Delta U = -(q^2 d)/(2 \varepsilon_\mathrm{e} s (s+d))$.

Now consider $\Delta F$ for opposite charges:
\begin{equation}
\label{ocdf}
\Delta F = 
\frac{q^2}{\varepsilon_\mathrm{e}} \int_0^\infty k e^{-2k(s+d)}
\left[ \frac{\alpha^2-1}{(1+\alpha e^{-2kd})^2} + 1 \right] \, \mathrm{d}k \ge 0.
\end{equation}
The magnitude of the attractive force between opposite charges with
a screening layer
is always less than when both charges are in one uniform dielectric medium with
dielectric constant $\varepsilon_\mathrm{e}$.  
This agrees with intuition upon letting $\varepsilon_\mathrm{e} \rightarrow 1$.
As expected, $\Delta F$ vanishes if $\alpha=0$.
Also, $\Delta F = q^2/(\varepsilon_\mathrm{e} (2s+2d)^2)$ if $\alpha=1$, which 
confirms that there is no force between charges that have a metal between them.
For $d=0$, the force difference 
$\Delta F = (q^2\alpha)/(2 \varepsilon_\mathrm{e} s^2 (1+\alpha))$
depends on $\alpha$ for the reason noted in the case of identical charges.

The energy difference when the comparison is made to the interaction with
the $\varepsilon_\mathrm{i}$ material everywhere is $\widetilde{\Delta U}$:
\begin{eqnarray}
\label{ocdut}
\widetilde{\Delta U} & = &  
\frac{q^2}{\varepsilon_\mathrm{e}} \int_0^\infty e^{-2k(s+d)}
\left[ \frac{\varepsilon_\mathrm{e}}{\varepsilon_\mathrm{i}} - 
\frac{1+\alpha e^{2kd}}{1+\alpha e^{-2kd}} \right] \, \mathrm{d}k \nonumber \\
 & = & \frac{q^2}{\varepsilon_\mathrm{e}\varepsilon_\mathrm{i}}
\int_0^\infty e^{-2k(s+d)}\left[ 
\frac{(\varepsilon_\mathrm{e}-\varepsilon_\mathrm{i}) 
+ \alpha(\varepsilon_\mathrm{e} e^{-2kd} - \varepsilon_\mathrm{i} e^{2kd})}
{1+\alpha e^{-2kd}} \right] \, \mathrm{d}k .
\end{eqnarray}
As expected on the basis of Fig.~\ref{U_d}), $\widetilde{\Delta U}$ 
is less than or equal to 0 
since both terms within the square brackets
are less than or equal to 0 in the case of screening ($0 \le \alpha \le 1$).
For $\alpha=0$, the energy difference $\widetilde{\Delta U}$ 
vanishes, while for $\alpha=1$ and $d \rightarrow \infty$, 
$\widetilde{\Delta U}=-q^2/(\varepsilon_\mathrm{e} 2s)$,
the energy of interaction due to the presence of image charges.
For $d=0$, 
$\widetilde{\Delta U}=q^2(\varepsilon_\mathrm{e}-\varepsilon_\mathrm{i})/
2\varepsilon_\mathrm{e}\varepsilon_\mathrm{i}s$.

Now consider $\widetilde{\Delta F}$:
\begin{eqnarray}
\label{ocdft}
\widetilde{\Delta F} & = & 
-\frac{q^2}{\varepsilon_\mathrm{i}} \int_0^\infty k e^{-2k(s+d)}
\left[ \frac{\varepsilon_\mathrm{i}(1-\alpha)(1+\alpha)}
{\varepsilon_\mathrm{e}(1+\alpha e^{-2kd})^2} - 1 \right] 
\, \mathrm{d}k \nonumber \\
& = & -\frac{q^2}{\varepsilon_\mathrm{i}} \int_0^\infty k e^{-2k(s+d)}
\left[ \frac{(1+\alpha)^2}{(1+\alpha e^{-2kd})^2} - 1 \right] \, \mathrm{d}k
\le 0 .
\end{eqnarray}
The attraction between unlike charges in our setting is 
always stronger than when the charges are in a uniform dielectric medium
of dielectric constant $\varepsilon_\mathrm{i}$.
Clearly, $\widetilde{\Delta F}$ vanishes when $\alpha=0$ and when $d=0$.

\section{Comments}

The energy and force for the case of two point charges in a dielectric medium
with a layer of differing dielectric between them has been compared with 
two baselines:  point charges in a uniform medium having the dielectric 
constant of the separating layer and point charges in a uniform medium 
having the dielectric constant of the exterior medium.
In the latter case,
we find that for opposite charges, $\Delta F > 0$ always, implying a weakened
attraction when compared to the baseline.
For identical charges, however, there are cases for which the repulsion is 
actually enhanced compared to this baseline.
Since it is possible to let $\varepsilon_\mathrm{e} \rightarrow 1$, this 
situation corresponds to an effective repulsion that is stronger than 
the vacuum case, a counter-intuitive result.
We refer to this behavior as `asymmetric screening'.

When both repulsion and attraction are weakened compared to
the $\varepsilon_\mathrm{e}$ baseline, which one is reduced more?  
This question is easily answered by considering
\[
\delta F \equiv \Delta F_{\mathrm{att}} - (-\Delta F_{\mathrm{rep}}) = 
\Delta F_{\mathrm{att}} + \Delta F_{\mathrm{rep}} .
\]
When $\delta F > 0$, there is a larger reduction of the attraction than
of the repulsion, and {\em vice versa}.
Using Eq.~(\ref{lcdf}) for $\Delta F_{\mathrm{rep}}$ 
and Eq.~(\ref{ocdf}) for $\Delta F_{\mathrm{att}}$,
we find
\[
\delta F = \Delta F_{\mathrm{att}} + \Delta F_{\mathrm{rep}} =
\frac{q^2}{\varepsilon_\mathrm{e}} \int_0^\infty k e^{-2k(s+d)}
\left[ \frac{1-\alpha^2}{(1-\alpha e^{-2kd})^2} - 
\frac{1-\alpha^2}{(1+\alpha e^{-2kd})^2} \right] \, \mathrm{d}k \ge 0 .
\]

For the case of the $\varepsilon_\mathrm{i}$ baseline, 
we see that $\widetilde{\Delta F}$ is always negative for opposite charges.
This indicates an enhanced attraction compared to the baseline 
(when both charges are in a uniform medium of dielectric constant
$\varepsilon_\mathrm{i}$).
For identical charges we have $\widetilde{\Delta F} > 0$, implying that the
repulsion is always enhanced when compared to this baseline.
One can consider 
\[
\widetilde{\delta F} \equiv \widetilde{\Delta F}_{\mathrm{att}} - 
(-\widetilde{\Delta F}_{\mathrm{rep}}) = 
\widetilde{\Delta F}_{\mathrm{att}} + \widetilde{\Delta F}_{\mathrm{rep}} .
\]
When $\widetilde{\delta F} > 0$, the repulsion of identical charges is enhanced
more then the attraction of opposite charges is.
Using Eq.~(\ref{lcdft}) for $\widetilde{\Delta F}_{\mathrm{rep}}$ 
and Eq.~(\ref{ocdft}) for $\widetilde{\Delta F}_{\mathrm{att}}$,
we find
\[
\widetilde{\delta F} = \widetilde{\Delta F}_{\mathrm{att}} + 
\widetilde{\Delta F}_{\mathrm{rep}} =
\frac{q^2}{\varepsilon_\mathrm{i}} \int_0^\infty k e^{-2k(s+d)}
\left[ \frac{(1+\alpha)^2}{(1-\alpha e^{-2kd})^2} - 
\frac{(1+\alpha)^2}{(1+\alpha e^{-2kd})^2} \right] \, \mathrm{d}k \ge 0 .
\]

According to Fig. \ref{F_d}, asymmetric screening is quite pronounced
at short ranges, and we expect the phenomenon to play an important role
in biomolecular recognition and in the adoption of the native conformation
of proteins.  
Particularly pronounced is the enhanced repulsion between charges of the 
same sign.
This behavior should exert a rather strong veto on poor matching of
charges as one part of a molecule interacts with another part or 
as two molecules interact with each other.
Therefore, accurate calculation of electrostatic interaction is essential
when considering biomolecular systems.  

\section*{Acknowledgements}
This research was supported by the Intramural
Research Program of the NIH, National Library of Medicine.

\appendix*
\section{Surface Charge Method}

The surface charge method\cite{tpd06}-\cite{tpd04} 
provides a relatively easy path to the induced surface charge.
In the case of two identical charges, symmetry implies that 
the induced surface charge densities on the two surfaces are
identical functions in the plane.
Therefore we may write
\begin{equation}
\label{scm:potgen}
\Phi = \frac{q}{\varepsilon_\mathrm{e}|\vec{r} - (d+s)\hat{z}|} +
\frac{q}{\varepsilon_\mathrm{e}|\vec{r} + (d+s)\hat{z}|} +
\int_{z'=+d} \frac{\sigma(\rho')}{|\vec{r}-\vec{r}'|} \, \mathrm{d}S' +
\int_{z'=-d} \frac{\sigma(\rho')}{|\vec{r}-\vec{r}'|} \, \mathrm{d}S'.
\end{equation}
The induced surface charge density $\sigma(\rho)$ is unknown, but can be
expanded in a complete set of functions.
Because of the cylindrical symmetry, 
Bessel functions are the obvious choice in this case.
Any reasonably well-behaved function $f(\rho)$ gives rise to the 
pair of transforms\cite{arfken1}
\begin{eqnarray*}
f(\rho) & = & \int_0^\infty a(\beta) J_\nu(\beta\rho) \, \mathrm{d}\beta\\
a(\beta) & = & \beta\int_0^\infty f(\rho) J_\nu(\beta\rho) \rho 
\, \mathrm{d}\rho,
\end{eqnarray*}
allowing us to write the surface charge as
\[
\sigma(\rho) = \int_0^\infty S(\beta) J_\nu(\beta\rho) \, \mathrm{d}\beta.
\]
Furthermore, the denominator of the integrals in Eq.~(\ref{scm:potgen}) 
can also be expanded in Bessel functions \cite{jdj}:
\[
\frac{1}{|\vec{r}-\vec{r}'|} = \sum_{m=-\infty}^{\infty} \int_0^\infty 
\mathrm{d}k \, e^{im(\phi-\phi')} J_m(k\rho) J_m(k\rho') e^{-k(z_>-z_<)},
\]
where $z_> = \max\{z,z'\}$ and $z_> = \min\{z,z'\}$.

In the vicinity of the surfaces, the potentials of the point charges are
\[
\frac{q}{\varepsilon_\mathrm{e}|\vec{r} - (d+s)\hat{z}|} = \frac{q}{\varepsilon_\mathrm{e}}
\int_0^\infty \mathrm{d}k \, J_0(k\rho) e^{-k(d+s-z)}
\]
and
\[
\frac{q}{\varepsilon_\mathrm{e}|\vec{r} + (d+s)\hat{z}|} = \frac{q}{\varepsilon_\mathrm{e}}
\int_0^\infty \mathrm{d}k \, J_0(k\rho) e^{-k(z+d+s)}.
\]

The potential near the boundary at $z=d$ due to the induced surface charge
at $z=d$ is
\begin{eqnarray*}
\int_{z'=+d} \frac{\sigma(\rho')}{|\vec{r}-\vec{r}'|} \, \mathrm{d}S' & = &
\int(\rho' \mathrm{d}\phi' \mathrm{d}\rho')
\left[\int_0^\infty {\cal S}(\beta) J_\nu(\beta\rho') 
\, \mathrm{d}\beta\right] \\
& & \times \left[ \sum_{m=-\infty}^\infty \int_0^\infty \mathrm{d}k \,
e^{im(\phi-\phi')} J_m(k\rho) J_m(k\rho') e^{-k(z_>-z_<)} \right] \\
& = & \int_0^\infty \mathrm{d}\beta\, {\cal S}(\beta) \int_0^\infty
\mathrm{d}k \, e^{-k(z_>-z_<)} \sum_{m=-\infty}^\infty J_m(k\rho)
\left[ \int \mathrm{d}\phi' e^{im(\phi-\phi')} \right] \\ & & \times
\left[ \int \mathrm{d}\rho' \rho' J_\nu(\beta\rho') J_m(k\rho') \right] \\
& = & 2\pi \int_0^\infty \mathrm{d}\beta\, {\cal S}(\beta) \int_0^\infty
\mathrm{d}k \, e^{-k(z_>-z_<)} J_0(k\rho) 
\left[ \int \mathrm{d}\rho' \rho' J_\nu(\beta\rho') J_0(k\rho') \right] .
\end{eqnarray*}
Letting $\nu = 0$ turns the $\rho'$ integral into a standard one,
\cite{arfken2}
\[
\int_0^\infty J_\nu(\beta\rho) J_\nu(\beta' \rho) \rho \, \mathrm{d}\rho = 
\frac{\delta(\beta-\beta')}{\beta} \qquad (v>-1/2),
\]
and therefore 
\[
\int_{z'=+d} \frac{\sigma(\rho')}{|\vec{r}-\vec{r}'|} \, \mathrm{d}S' =
2\pi \int_0^\infty \mathrm{d}k \, e^{-k(z_>-z_<)} J_0(k\rho) {\cal S}(k)/k .
\]
So for $z'=+d$ and $z>d$ (just above the top interface)
\[
2\pi \int_0^\infty \mathrm{d}k \, e^{-k(z-d)} J_0(k\rho) {\cal S}(k)/k .
\]
For $z'=+d$ and $z<d$ (just below the top interface)
\[
2\pi \int_0^\infty \mathrm{d}k \, e^{-k(d-z)} J_0(k\rho) {\cal S}(k)/k .
\]
For $z'=-d$ and $z$ near $d$ one finds a similar formula that is valid either 
above or below interface:
\[
2\pi \int_0^\infty \mathrm{d}k \, e^{-k(z+d)} J_0(k\rho) {\cal S}(k)/k .
\]

The boundary condition at $z=d$ is
\[
\varepsilon_\mathrm{i} \left. \frac{\partial \Phi_{z \le d}}{\partial z} \right |_{z=d} =
\varepsilon_\mathrm{e} \left. \frac{\partial \Phi_{z \ge d}}{\partial z} \right |_{z=d}
\]
for every value of $\rho$, which leads to an equation easily solved for 
${\cal S}(k)$:
\[
{\cal S}(k) = \frac{q k \alpha e^{-ks}(e^{-2kd}-1)}
{2\pi\varepsilon_\mathrm{e}(1-\alpha e^{-2kd})} .
\]
Therefore
\begin{eqnarray*}
\sigma(\rho) & = & \int_0^\infty J_0(k\rho) {\cal S}(k) \, \mathrm{d}k \\
& = & \frac{q \alpha}{2 \pi \varepsilon_\mathrm{e}} \int_0^\infty J_0(k\rho)
\frac{k e^{-ks}(e^{-2kd}-1)}{(1-\alpha e^{-2kd})} \, \mathrm{d}k \\
& = & -\frac{q \alpha}{2 \pi \varepsilon_\mathrm{e}} \int_0^\infty J_0(k\rho)
k e^{-ks} \left(1+\frac{(\alpha-1)e^{-2kd}}{(1-\alpha e^{-2kd})}\right) \, \mathrm{d}k \\
& = & -\frac{q \alpha}{2 \pi \varepsilon_\mathrm{e}} \left[ 
\int_0^\infty J_0(k\rho) k e^{-ks} \, \mathrm{d}k +
\int_0^\infty J_0(k\rho) k e^{-k(s+2d)} (\alpha-1) \sum_{n=0}^\infty 
\alpha^n e^{-2knd} \, \mathrm{d}k \right] \\
& = & -\frac{q \alpha}{2 \pi \varepsilon_\mathrm{e}} \left[ 
\int_0^\infty J_0(k\rho) k e^{-ks} \, \mathrm{d}k +
\sum_{n=0}^\infty \alpha^n (\alpha-1) \int_0^\infty J_0(k\rho) k 
e^{-k(s+2(n+1)d)} \, \mathrm{d}k \right] .
\end{eqnarray*}
Since all variables are real, we make use of the following integral\cite{GR1}
\[
\int_0^\infty e^{-\alpha x} J_\nu(\beta x) x^{\nu+1} \, \mathrm{d}x = 
\frac{(2\alpha)(2\beta)^\nu \Gamma(\nu+(3/2))}
{\sqrt{\pi}(\alpha^2+\beta^2)^{\nu+(3/2)}}
\]
for $\nu>-1$ and $\alpha>0$.
Recall that $\Gamma(n+(1/2)) = \sqrt{\pi}(2n-1)!!2^{-n}$, 
so that $\Gamma(3/2)=\sqrt{\pi}/2$.
Therefore, the surface charge density is
\[
\sigma(\rho) = -\frac{q \alpha}{2 \pi \varepsilon_\mathrm{e}} \left[
\frac{s}{(s^2+\rho^2)^{3/2}} + \sum_{n=0}^\infty \alpha^n (\alpha-1)
\frac{s+2(n+1)d}{\left[ (s+2(n+1)d)^2+\rho^2 \right]^{3/2}}
\right] ,
\]
from which it is easy to verify that 
$\int \sigma(\rho) 2\pi\rho \mathrm{d}\rho = 0$.

This charge density can be used to recover same energy and force as before.
To compute the energy (and then the force), $\Phi(\rho=0, z=s+d)$ must be
computed from $\sigma(\rho)$.
For $z'=d$, $\rho=0$, and $z=s+d$, $|\vec{r}-\vec{r}'|^2=s^2+\rho'^2$.
Therefore
\begin{eqnarray*}
\int \frac{\sigma(\rho')}{|\vec{r}-\vec{r}'|} \, \mathrm{d}S' & = &
2 \pi \int \frac{\rho'\sigma(\rho')}{(s^2+\rho'^2)^{1/2}} \, \mathrm{d}\rho' \\
& = & 2 \pi \int \left( {\cal S}(k) 
\int \frac{\rho'J_0(k\rho')}{(s^2+\rho'^2)^{1/2}} \, \mathrm{d}\rho'
\right) \, \mathrm{d}k .
\end{eqnarray*}
The $\rho'$ integral is found in tables\cite{GR2} to be
\[
\int_0^\infty \frac{xJ_0(xy)}{(a^2+x^2)^{1/2}} \,\mathrm{d}x = 
\frac{e^{-ay}}{y} ,
\]
and so
\[
\int \frac{\sigma(\rho')}{|\vec{r}-\vec{r}'|} \, \mathrm{d}S'  = 
\frac{q \alpha}{\varepsilon_\mathrm{e}} \int_0^\infty e^{-2ks}
\frac{e^{-2kd}-1}{1-\alpha e^{-2kd}} \, \mathrm{d}k .
\]
For $z'=-d$, $\rho=0$, and $z=s+d$, $|\vec{r}-\vec{r}'|^2=(s+2d)^2+\rho'^2$.
The contribution to the potential from the induced surface charge at $z'=-d$ is
\[
\int \frac{\sigma(\rho')}{|\vec{r}-\vec{r}'|} \, \mathrm{d}S'  = 
\frac{q \alpha}{\varepsilon_\mathrm{e}} \int_0^\infty e^{-2k(s+d)}
\frac{e^{-2kd}-1}{1-\alpha e^{-2kd}} \, \mathrm{d}k .
\]
The potential at $\rho=0$ and $z=s+d$ is
\begin{eqnarray*}
\Phi(\rho=0,z=s+d) & = & \frac{q}{\varepsilon_\mathrm{e}} \int_0^\infty \! \left(
e^{-k(2s+2d)} + \frac{\alpha e^{-2ks} (e^{-2kd}-1)}{1-\alpha e^{-2kd}} + 
\frac{\alpha e^{-k(2s+2d)} (e^{-2kd}-1)}{1-\alpha e^{-2kd}}
\right) \mathrm{d}k \\
& = & \frac{q}{\varepsilon_\mathrm{e}} \int_0^\infty e^{-2k(s+d)} 
\frac{1-\alpha e^{2kd}}{1-\alpha e^{-2kd}} \, \mathrm{d}k ,
\end{eqnarray*}
and therefore
\[
U = \frac{q^2}{\varepsilon_\mathrm{e}} \int_0^\infty e^{-2k(s+d)}
\frac{1-\alpha e^{2kd}}{1-\alpha e^{-2kd}} \, \mathrm{d}k ,
\]
in agreement with Section \ref{lc}.  
Because $U$ agrees, everything that follows from $U$ must also agree.

For opposite charges
\[
\Phi = \frac{q}{\varepsilon_\mathrm{e}|\vec{r} - (d+s)\hat{z}|} -
\frac{q}{\varepsilon_\mathrm{e}|\vec{r} + (d+s)\hat{z}|} +
\int_{z'=+d} \frac{\sigma_+(\rho')}{|\vec{r}-\vec{r}'|} \, \mathrm{d}S' +
\int_{z'=-d} \frac{\sigma_-(\rho')}{|\vec{r}-\vec{r}'|} \, \mathrm{d}S'
\]
However, by symmetry $\sigma_+=-\sigma_-\equiv \sigma$.
The boundary condition yields
\[
{\cal S}(k) = \frac{q k \alpha e^{-ks}(e^{-2kd}+1)}
{2\pi\varepsilon_\mathrm{e}(1+\alpha e^{-2kd})}
\]
The surface charge density becomes
\[
\sigma(\rho) = \frac{q \alpha}{2 \pi \varepsilon_\mathrm{e}} \left[
\frac{s}{(s^2+\rho^2)^{3/2}} + \sum_{n=0}^\infty (-\alpha)^n (1-\alpha)
\frac{s+2(n+1)d}{\left[ (s+2(n+1)d)^2+\rho^2 \right]^{3/2}}
\right]
\]
Again, it is easy to verify that $\int \sigma(\rho) 2\pi\rho \mathrm{d}\rho = 0$
and that the energy $U$ reproduces the result in Section \ref{oc}.

\end{document}